\documentstyle[a4,12pt,epsf]{article}

\newcommand{\tr}{{\rm Tr}}

\newcommand{\la}{\langle}
\newcommand{\ra}{\rangle}
\newcommand{\be}{\begin{equation}}
\newcommand{\ee}{\end{equation}}
\newcommand{\bea}{\begin{eqnarray}}
\newcommand{\eea}{\end{eqnarray}}

\newcommand{\pa}{\partial}

\newcommand{\cd}{{\cal D}}
\newcommand{\nn}{\nonumber \\}
\newcommand{\f}{\frac}
\newcommand{\cl}{{\cal L}}
\newcommand{\p}{\phi}
\newcommand{\ppc}{\phi_c}
\newcommand{\lm}{\lambda}
\newcommand{\lqn}{\lefteqn}

\newcommand{\intk}{\int \f{d^4k}{(2\pi)^4}}

\newcommand{\intbk}{{\textstyle \sum} \hspace{-1em} \int_{\beta,k}}
\newcommand{\fpi}{f_\pi}

\begin{document}
\begin{center}
{\Large \bf Effective Potential of $\bf O(N)$ Linear Sigma Model}\\
{\Large \bf at Finite Temperature}\\
%
\vskip 10mm
Y. Nemoto$^{*,**}$\footnote{E-mail address: nemoto@rcnp.osaka-u.ac.jp}
, K. Naito$^{*,***}$ and M. Oka$^*$\\
\vspace{0.5em}
{\it $^*$Department of Physics, Tokyo Institute of Technology,} \\
{\it Meguro, Tokyo 152-8551 Japan}\\
\vspace{0.5em}
{\it $^{**}$Research Center for Nuclear Physics (RCNP), Osaka University,}\\
{\it Ibaraki, Osaka 567-0047 Japan}\footnote{Present address}\\
\vspace{0.5em}
{\it $^{***}$Radiation Laboratory, the Institute of Physical and 
Chemical Research (RIKEN)}\\
{\it Wako, Saitama 351-0198 Japan}\footnote{Present address}\\
\end{center}
\vspace{2em}
\begin{abstract}
\baselineskip 1.2em
  We study the $O(N)$ symmetric 
linear sigma model at finite temperature as the low-energy
effective models of quantum chromodynamics(QCD) using the 
Cornwall-Jackiw-Tomboulis(CJT) effective action for composite operators.
It has so far been claimed that the Nambu-Goldstone theorem is not
satisfied at finite temperature in this framework unless the
large $N$ limit in the $O(N)$ symmetry is taken.
We show that this is not the case.
The pion is always massless below the critical temperature,
if one determines the propagator within the form such that the
symmetry of the system is conserved,
and defines the pion mass as the curvature of the effective potential.
We use a new renormalization prescription for the CJT effective
potential in the Hartree-Fock approximation.
A numerical study of the Schwinger-Dyson equation and the
gap equation is carried out including the thermal and quantum loops.
We point out a problem in the derivation of the sigma meson mass without
quantum correction at finite temperature.
A problem about the order of the phase transition in this approach 
is also discussed.

\end{abstract}

\clearpage

\section{Introduction}  
\baselineskip 1.2em

  Chiral symmetry is one of the most important features of low-lying 
hadron properties.
In quantum chromodynamics(QCD) this symmetry is well satisfied in the
$SU(2)$ sector due to the light $u,d$ quark masses.
Existence of the nearly massless mesons, or pions, means that the chiral
symmetry must be spontaneously broken and the Nambu-Goldstone(NG)
bosons appear.
Spontaneous chiral symmetry breaking(SCSB) is also manifested in the
non-degenerate parity doublet of nucleons.
So SCSB influences the low-lying hadron spectra significantly.
Though such low-energy phenomena of the strong interaction as well as
the quark confinement are now widely
confirmed in experiments, no analytic investigation based on the first
principle, or QCD is carried out due to its non-perturbative nature.
So it is still important to study the effective models of QCD in order to
understand the nature of hadrons in addition
to numerical analyses in the lattice gauge theory.

  In increasing temperature, as shown in the lattice QCD calculation, 
the chiral symmetry is believed to be restored at around $T=100\sim 300$ MeV.
The study of physics at finite temperature is very interesting from both
theoretical and experimental points of view.
The big bang model tells us that a series of phase
transitions, which of course include the QCD phase transition, occurred 
in the early universe.
It could also be possible to probe the 
underlying physics of QCD in laboratory involving
relativistic heavy ion collisions. 
These experiments are planned in near future 
and its result is expected to elucidate some important questions 
like the mechanism of the chiral symmetry restoration and
the nature of quark gluon plasma.

  Theoretical investigation of the symmetry restoration at finite
temperature in terms of field theory was first studied by 
Kirzhnits and Linde\cite{Kir72}.
They observed that spontaneous symmetry breaking(SSB) will be restored at
sufficiently high temperatures.
It is now well-known that the naive effective potential up to the 1-loop
level does not work at finite temperature.
Weinberg pointed out that
at very high temperature, powers of temperature $T$ can compensate for
powers of a small coupling constant $e$, leading to a breakdown of the
perturbation expansion\cite{Wei74}.
He showed that the leading effect of this sort arises from the $e^2T^2$
term in the scalar $\p^4$ theory.
Dolan and Jackiw and others showed that systematic summation of a certain 
kind of loop diagrams in this model is needed at finite temperature
\cite{Dol74,Kir76}.

  For analysis of the linear sigma model at finite temperature, 
the mean field approximation has mainly been used so far.
The mean field theory is a theory in which the resummation of the diagrams
is introduced approximately by hand and has been used by many authors
\cite{Bay77,Lar86,Roh98}.

  Recently several analyses at finite temperature have been done using an 
extended
effective potential for composite operators introduced by Cornwall, Jackiw, 
and Tomboulis(CJT) \cite{CJT74}.
Contrary to the usual effective action, their effective action depends
not only the classical field $\phi_c(x)$ but also on $G(x,y)$.
These two quantities are to be realized as the expectation values 
of a quantum field  $\phi(x)$ and the
time ordered product of the field operator $T\phi(x)\phi(y)$ 
respectively.
In this case the effective action $\Gamma(\phi_c,G)$ is the generating 
functional of the two particle irreducible vacuum graphs.
This formalism was originally written at zero
temperature but it has been extended for finite temperature in $\phi^4$
theory by Amelino-Camelia and Pi\cite{Ame93}.

  There is an advantage in using the CJT formalism to calculate the 
effective potential in the Hartree-Fock approximation.  
According to ref.\cite{Ame93},
we need to evaluate only one graph that of the ``double bubble''
instead of summing infinite many ``daisy'' and 
``super daisy'' graphs
using the usual tree level propagators.
These infinite diagrams are incorporated automatically in the effective
action if the CJT formalism is used,
so the effective potential takes real values at finite temperature.
The need of the resummation of this kind of loop diagrams at
finite temperature is also discussed in other 
approaches \cite{Chi98}.

  Amelino-Camelia {\it et.al.} investigated the $O(N)$ linear sigma model,
which is regarded as the low-energy effective model of the sigma meson and
the pions,
at finite temperature with the CJT action\cite{Ame97,Pet98}.
It is known, however, that in their approach one encounters a difficulty
that the NG theorem appears to be violated in the broken symmetry
phase at finite temperature.
They concluded that in the CJT formalism the NG theorem
is violated in the Hartree-Fock approximation, i.e., 
the case of finite values of $N$ and the so-called super-daisy
approximation,
while it is satisfied if and only if leading contributions in the $1/N$ 
expansion are taken.
They also discussed the renormalization in the CJT formalism and suggested that
a consistent renormalization cannot be performed in the broken phase
since the non-perturbative quantities, or solutions of the 
Schwinger-Dyson(SD) equations, are included.

 In this paper we study two subjects.
One is the formulation of the linear sigma model at finite temperature.
We re-analyze the $O(N)$ linear sigma model at finite temperature
in the CJT formalism and show that the NG theorem is always satisfied
at any finite values of $N$ in the Hartree-Fock approximation.
The CJT action does not violate the $O(N)$ symmetry, so that
the pions are always massless below the critical temperature
if we define the meson masses by the curvature, or the second derivative of 
the effective potential.
On the renormalization, we adopt the method of the renormalization of
auxiliary fields.
As shown in the CJT's paper, the solutions of the SD-equation can be thought
as a kind of auxiliary field in the mean field level.
We show that the CJT effective potential is renormalizable if the technique of
the auxiliary field is applied.
This approach has a clear advantage that once the effective potential is
renormalized, all the physical quantities derived from the effective
potential are finite.

The other is the application of the above formalism to low-energy hadron
properties.
The linear sigma model has various merits as an effective model of
low-energy hadron dynamics.
It can describe SSB as in QCD.
Furthermore one can study both the symmetry broken phase and the restored
phase if desired.
Other effective models such as non-linear sigma model or chiral perturbation
theory treat basically the world with SSB.

The contents of this paper are as follows.
In Sec. \ref{chp:cjt}, we introduce the CJT effective action for composite
operators and explain the difference from the ordinary effective
action.
In Sec. \ref{chp:on}, we construct an effective potential for the
$O(N)$ symmetric linear sigma model.
Here we also point out defects of the conventional formulation.
We show that the NG theorem holds even at finite temperature
if one adopts the $O(N)$ symmetric form of the propagator and defines 
the meson masses
as the curvatures of the effective potential.
We also give a new renormalization prescription of the CJT effective
potential.
It is based on the renormalization of auxiliary fields and was long ago
applied to this model in the large $N$ limit\cite{CJP74,Kob75}.
It can remove divergences in the effective potential both in the
symmetry broken and unbroken phases, while
the conventional renormalization can be performed only in the restored 
phase.
Using these techniques, we apply the linear sigma model to the low-energy
mesons in Sec. \ref{chp:app1}.
As is well-known, the $O(4)$ linear sigma model can describe the physics of
the sigma meson and the pions in low-energy region.
We discuss temperature dependence of physical masses of these mesons and
sigma condensate corresponding to the pion decay constant.
In Sec. \ref{sec:comm}, we concentrate on the order of the phase transition and
consider the so-called setting-sun diagram.
We discuss approximate treatments of the setting-sun diagram and compare
them.
Finally summary and conclusion are given in Sec. \ref{chp:summ}.

\section{Effective Action for Composite Operators}
    \label{chp:cjt}

  Effective action for composite operators was originally studied in 
condensed matter physics \cite{Lee60}
 and then extended to relativistic field theories
\cite{Dah67}.
We here use an approach introduced by Cornwall, Jackiw and Tomboulis
\cite{CJT74}.
This is based on functional methods or the path integral representation
for the Green functions.

  The effective action for composite operators is a generalization of 
the conventional effective action and is written by $\Gamma[\ppc,G]$.
This is a functional both of the expectations value of the quantum
field $\ppc(x)=\la 0|\p(x)|0 \ra$ and of the propagator 
$G(x,y) = \la 0|{\rm T}\p(x)\p(y) |0\ra$.
The $c$-number function $\ppc(x)$ is also called a classical field.
The variational equations
\be
  \f{\delta \Gamma[\ppc,G]}{\delta \ppc(x)} = 0,
  \label{eq:cjt-gap}
\ee
\be
  \f{\delta \Gamma[\ppc,G]}{\delta G(x,y)} = 0
  \label{eq:cjt-sd}
\ee
determine $\ppc$ and $G$ on the vacuum.
The equation (\ref{eq:cjt-sd}) is nothing but the SD-equation for the
propagator $G$.
Furthermore, we can show that the second variational derivative of
$\Gamma[\ppc,G]$ in $G$ leads to the Bethe-Salpeter equation which
describes relativistic bound states.
For example, in hadron physics,
this is used in order to describe meson states as the bound state of
the quark-antiquark pair.

  We now describe the series expansion for $\Gamma[\p,G]$.
We introduce the so-called tree level 2-point Green function by
\be
  i{\cal D}^{-1}(\ppc,x,y) 
  = \f{\delta^2 S[\ppc]}{\delta \ppc(x) \delta \ppc(y)} 
  = iD^{-1}(x-y) + \f{\delta^2 S[\ppc]_{\rm int}}{\delta \ppc(x) 
  \delta \ppc(y)}
\ee
The required series is then
\be
  \Gamma[\ppc,G] = S[\ppc] + \f{i}{2} \tr {\rm Ln} G^{-1} 
  + \f{i}{2} \tr \cd^{-1}(\ppc) G + \Gamma_2[\ppc,G] +\rm const
  \label{eq:cjt-cjt}
\ee
where the trace, the logarithm and the product $\cd^{-1}G$ are taken
in the functional sense.
The conventional effective action $\Gamma[\ppc]$ is $\Gamma[\ppc,G]$ with $I=0$,
i.e.,
\be
  \Gamma[\ppc] = \Gamma[\ppc,G_0]
  \label{eq:cjt-convgamma1}
\ee
\be
  \f{\delta \Gamma[\ppc, G_0]}{\delta G_0(x,y)}=0
  \label{eq:cjt-convgamma2}
\ee
The constant which is independent of $\p$ and $G$ is evaluated so that
eqs.(\ref{eq:cjt-convgamma1}) and (\ref{eq:cjt-convgamma2}) are satisfied:
\be
  \Gamma[\ppc,G] = S[\ppc] + \f{i}{2} \tr {\rm Ln} DG^{-1}
  + \f{i}{2} \tr (\cd^{-1}(\ppc) G-1) + \Gamma_2[\ppc,G]
\ee
This expression is used in some literature.
$\Gamma_2[\ppc,G]$ is given by all the two-particle and higher two-particle
irreducible vacuum graphs in a theory which has the vertices determined by
the interaction of the action 
$S_{\rm int}[\p,\ppc]$ and the propagators $G(x,y)$.

  When one considers the case of translation-invariant solutions,
one sets $\ppc(x)$ to a constant $\ppc$ and takes $G(x,y)$ to be a
function only of $x-y$.
The series for the effective potential $V(\ppc,G)$ can be easily obtained 
from eq.(\ref{eq:cjt-cjt}):
\be
  V(\ppc,G) = V_0(\ppc) - \f{i}{2}\intk \ln G^{-1}(k) 
  -\f{i}{2}\intk {\rm tr}\cd^{-1}(\ppc,k)G(k) + V_2(\ppc,G) 
  \label{eq:cjt-veff}
\ee
where $V_0(\ppc)$ is the tree-level (classical) effective potential,
\be
  G(k) = \int d^4x e^{ik\cdot (x-y)}G(x-y)
\ee
\be
  \cd(\ppc,k) = \int d^4x e^{ik\cdot (x-y)}\cd(\ppc,x-y)
\ee
and $-V_2(\ppc,G)$ is the sum all two and higher order loop two-particle
irreducible vacuum graphs of the theory with vertices given by 
$S_{\rm int}(\ppc)$ and propagator $G(k)$.
The stationary requirements are then
\be
  \f{\pa V(\ppc,G)}{\pa \ppc} = 0
  \label{cjt:eq-gapcon}
\ee
\be
  \f{\pa V(\ppc,G)}{\pa G(k)} = 0
\ee
These equations are the starting point of our discussion.

\section{Formulation of the $O(N)$ Linear Sigma Model} 
    \label{chp:on}                                     

  In this section we construct the effective potential of the $O(N)$ 
symmetric linear
sigma model using the CJT formalism.
In QCD with two-flavor massless quarks, the chiral $SU(2)_L\times SU(2)_R$
symmetry is satisfied in the lagrangian level.
But this symmetry is spontaneously broken to the $SU(2)_V$ symmetry,
because the vacuum, i.e., the ground state of the field configuration violates
it.
We can see this phenomenon in the $O(4)$ linear sigma model,
because the $O(4)$ symmetry has the same algebra as $SU(2)\times SU(2)$.
Four fields in the $O(4)$ linear sigma model are identified with
one sigma meson and three pions when we regard it as an effective model
of hadrons.
By spontaneous symmetry breaking from $O(4)$ to $O(3)$, the three fields
become massless according to the NG theorem and the remaining one is massive.
Here we concentrate on the meson fields and see how this model is formulated
with the CJT effective action.
We set the arbitrary $N$ instead of $N=4$ at the stage of the formulation
for generality.
We use the Euclidean metric hereafter because it is convenient when we extend
the model to that with the imaginary time formalism 
at finite temperature.

\begin{figure}[t]
  \hspace*{3cm}
  \epsfxsize=250pt
  \epsfbox{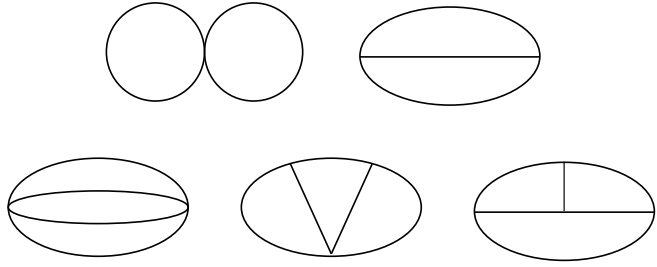}
  \caption{Two-particle irreducible diagrams contributing to $\Gamma_2[\ppc,G]$
  up to the three-loop level.}
  \label{fig:irred}
\end{figure}

  The lagrangian density for the $O(N)$ linear sigma model is given by
\be      \label{eq:lag1}
  \cl = \f{1}{2}\pa_\mu \p^a \pa_\mu \p^a +\f{1}{2} m^2 \p^2
        +\f{\lm}{6N}(\p^2)^2
\ee
where $\p^2=\p^a\p^a$, $a$ runs over 1 to $N$ and
repeated indices are summed.
By shifting the field as $\p^a(x) = \p^a(x) + \p^a_c(x)$,
The tree-level 2-point Green function $\cd$ is obtained:
\bea
  \cd_{ab}^{-1}(\p_c,x,y) 
  &=& \left. \f{\delta^2 S}{\delta \p^a \delta \p^b} \right|_{\p=\ppc} \nn
  &=& \left[-\pa_\mu\pa_\mu +m^2+\f{2\lm}{3N}\p^2_c\right] 
  \delta^{ab}\delta^4(x-y)
  +\f{4\lm}{3N}\p^a_c \p^b_c \delta^4(x-y)
\eea
In momentum space, it is given by
\be
  \cd_{ab}^{-1}(\p_c,k) =
  \left[k^2 +m^2+\f{2\lm}{3N}\p^2_c\right]
  \delta^{ab}
  +\f{4\lm}{3N}\p^a_c \p^b_c
  \label{eq:cdk-1}
\ee
and the inverse becomes
\be
  \cd_{ab}(\p_c,k) = \f{1}{k^2 +m^2+\f{2\lm}{N}\p^2_c}
  \f{\ppc^a\ppc^b}{\ppc^2}
  +\f{1}{k^2 +m^2+\f{2\lm}{3N}\p^2_c}
  \left( \delta^{ab}-\f{\ppc^a\ppc^b}{\ppc^2}\right)
\ee
The interaction lagrangian which describes the vertices of the shifted
theory is given by
\be
  \cl_{\rm int}(\ppc,\p) = \f{2\lm}{3N}\p^2 \p^a_c \p^a 
  +\f{\lm}{6N}(\p^2)^2
\ee
The diagrams contributing to $\Gamma_2[\ppc,G]$ are shown in 
Fig. \ref{fig:irred}.
Each line represents the propagator $G_{ab}(x,y)$,
and there are two kinds of vertices:
a four-point vertex proportional to $\lm$ and a three-point vertex, which
results from shifting the fields, proportional to $\lm\ppc^a(x)$.

\begin{figure}[h]
  \hspace*{3cm}
  \epsfxsize=300pt
  \epsfbox{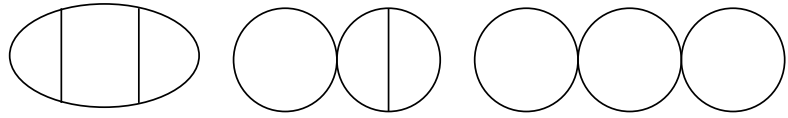}
  \caption{Two-particle reducible diagrams which do not contribute to 
  $\Gamma_2[\ppc,G]$ up to the three-loop level.}
  \label{fig:red}
\end{figure}

If we were computing the ordinary effective action $\Gamma[\ppc]$,
the lines would represent the propagator $\cd_{ab}(\ppc,x,y)$
and there would be additional contributions which are two-particle
reducible.
These diagrams are shown in Fig. \ref{fig:red}.

  Let us compute the CJT effective action(\ref{eq:cjt-cjt}).
We evaluate $\Gamma[\ppc.G]$ in the Hartree-Fock approximation, i.e.,
we take into account the ``$\infty$" type diagram only.
This diagram is a leading order in $\Gamma_2[\ppc,G]$ 
in both the loop expansion and the $1/N$ expansion.
The other two-loop diagram, the setting-sun type diagram, which is the next to
leading order in the $1/N$ expansion, is quite lengthy
to calculate and causes a problem on renormalization (see later).

In this approximation, $\Gamma_2$ is given by
\be
  \Gamma_2(\p_c,G) = \f{\lm}{6N} \int d^4x ( G_{aa}(x,x)G_{bb}(x,x)+
  2G_{ab}(x,x)G_{ba}(x,x) )
\ee
Then the CJT action has the form
\bea
 \Gamma[\p_c,G] &=& S[\p_c] + \f{1}{2} \tr {\rm Ln} G^{-1}
  + \f{1}{2} \tr (\cd^{-1}(\p_c)G) \nn
 && +\f{\lm}{6N} \int d^4x ( G_{aa}(x,x)G_{bb}(x,x)+
  2G_{ab}(x,x)G_{ba}(x,x) )
  \label{eq:CJT2}
\eea
where $S[\ppc]$ is the classical action
\be
  S[\p_c] = \int d^4x \left( \f{1}{2}\pa_\mu \p^a_c \pa_\mu \p^a_c
  +\f{1}{2}m^2 \ppc^2 +\f{\lm}{6N}(\ppc^2)^2 \right)
\ee
From this, the effective potential for the composite operators is given by
\bea
  \lefteqn{V(\ppc,G)}\nn
  &=& V_0[\p_c] + \f{1}{2} \int \f{d^4k}{(2\pi)^4}\ln \det G^{-1}(k)
  + \f{1}{2} \int \f{d^4k}{(2\pi)^4}{\rm tr} (\cd^{-1}(\p_c,k)G(k)) \nn
 && +\f{\lm}{6N} \left[ \left\{ \int \f{d^4k}{(2\pi)^4}  G_{aa}(k)\right\}^2+
  2\left\{\int \f{d^4k}{(2\pi)^4} G_{ab}(k)\right\}
  \left\{\int \f{d^4k}{(2\pi)^4} G_{ba}(k)\right\}\right]
\eea
with
\be
  V_0[\ppc] = \f{1}{2}m^2 \ppc^2 + \f{\lm}{6N}(\ppc^2)^2
\ee
where the trace and determinant are taken only in the internal (flavor)
space.

 If one applies this model to a system at finite temperature,
one replaces the loop integral with
\be
  \int \f{d^4k}{(2\pi)^4} \longrightarrow \f{1}{\beta}\sum_{n=-\infty}^\infty
  \int \f{d^3k}{(2\pi)^3} \equiv \intbk
\ee
in the imaginary time formalism.
Here $\beta\equiv 1/T$ is the inverse temperature and the sum is taken over
the Matsubara frequency.
So the starting effective potential we study is
\bea
  V(\ppc,G)
  &=& V_0[\p_c] + \f{1}{2} \intbk \ln \det G^{-1}(k)
  + \f{1}{2} \intbk (\cd^{-1}(\p_c,k)G(k)) \nn
 && +\f{\lm}{6N} \left[ \left\{ \intbk  G_{aa}(k)\right\}^2+
  2\left\{\intbk G_{ab}(k)\right\}
  \left\{\intbk G_{ba}(k)\right\}\right]
\eea

Minimizing the effective potential with respect to the
propagator $G(\ppc,k)$, we obtain the SD-equation
\be    \label{eq:sd1}
  G_{ab}^{-1}(p) = \cd_{ab}^{-1}(\ppc,p)
  +\f{2\lm}{3N}\left[ \intbk G_{cc}(k)\delta^{ab} + 
  2\intbk G_{ab}(k)\right]
\ee
The solution $G_0(\p_c,k)$ of this equation is inserted back
into the expression for the effective potential to give the 
conventional effective potential as a function of $\ppc$. 

\subsection{How to Take the Form of the Propagator $G$}  

\subsubsection{Conventional Forms of the Propagator $G$} 

  First we need to determine the form of the SD-equation, $G$.
In case of the $O(N)$ linear sigma model,
it seems to be natural to adopt the following multi-component form,
\be  \label{eq:dress}
  G(k)_{ab}(\ppc,k) = \delta^{a1}\delta^{b1}\f{1}{k^2+M_\sigma^2}
  +\sum_{j=2}^{N} \delta^{aj}\delta^{bj}\f{1}{k^2+M_\pi^2} 
\ee
In fact several authors adopt this form\cite{Ame97,Pet98}
and in ref.\cite{Pet98} this is called "dressed propagator" ansatz.
They also identify the parameters $M_\sigma$ and $M_\pi$ with the masses
of the sigma meson and pions, respectively.
It is, however, obvious that this form of $G$ does violate the $O(N)$
symmetry.
So the effective potential after substituting the solution $G_0$ also
violates the $O(N)$ symmetry.
This is one of the reasons why this formulation erroneously concludes that
the pions which should be massless below
the critical temperature acquire finite masses.

\subsubsection{An $O(N)$ Symmetric Form of $G$}
   \label{subsec:ours}

  We calculate the meson masses by using the definitions 
(\ref{eq:msigma-def}) and (\ref{eq:mpi-def}) from the effective potential
directly.
In the CJT formalism, we need to choose the form of $G$ not to break the
symmetry.
If one takes eq.(\ref{eq:dress}), for instance, we will lose the $O(N)$
symmetry in the resulting effective potential.
Here we use the following form of $G$ from the analogy of $\cd$.
\be        \label{eq:Ginv}
  G^{-1}_{ab}(\ppc,k) = (k^2+m^2+\f{X(\ppc^2)}{N}\ppc^2)\delta^{ab}
  +\f{Y(\ppc^2)}{N}\ppc^a\ppc^b
\ee
where $X(\ppc^2)$ and $Y(\ppc^2)$ are unknown functions which are
determined by solving the SD-equation.
We assume that $X$ and $Y$ are independent of the external momentum and 
also that no more parameters are needed because
the Hartree-Fock approximation is employed.

Substituting eq.(\ref{eq:Ginv}) into the CJT effective 
potential(\ref{eq:cjt-veff}), we obtain
\bea   \label{eq:CJTeff-2}
  \lefteqn{V(\ppc,G(\ppc,k))}\nn
  &=& \f{m^2}{2}\ppc^2+\f{\lm}{6N}(\ppc^2)^2
  + \f{1}{2}Q[M_{XY}] +\f{N-1}{2}Q[M_{X}] \nn
  &&+\f{1}{2} \left(m^2+\f{2\lm}{N}\ppc^2-M_{XY}^2\right)
  P[M_{XY}] 
  +\f{N-1}{2}\left(m^2+\f{2\lm}{3N}\ppc^2-M_{X}^2\right)
  P[M_{X}] \nn
  &&+\f{\lm}{6N}\left[ 3P[M_{XY}]^2 
  + (N^2-1)P[M_{X}]^2
  +2(N-1)P[M_{XY}]P[M_{X}]\right]
\eea
where we introduce
\be
  M_{X}^2 \equiv m^2 + \f{X(\ppc^2)}{N}\ppc^2
\ee
\be
  M_{XY}^2 \equiv m^2 + \f{X(\ppc^2)+Y(\ppc^2)}{N}\ppc^2
\ee
and
\be
  P[m] = \intbk \f{1}{k^2+m^2}
  \label{eq:on-P}
\ee
\be
  Q[m] = \intbk \ln (k^2+m^2)
\ee
for simplicity.
This effective potential depends only on $\ppc^2\equiv\ppc^a\ppc^a$
and therefore is $O(N)$ symmetric obviously.

The functions $M_{X}$ and $M_{XY}$ are determined from the following 
SD-equations.
\be
  M_{XY}^2
  = m^2+\f{2\lm}{N}\ppc^2+\f{2\lm}{N}
  P[M_{XY}]
  +\f{2\lm}{3}\f{N-1}{N}P[M_{X}]
  \label{eq:sds}
\ee
\be
  M_{X}^2
  = m^2+\f{2\lm}{3N}\ppc^2+\f{2\lm}{3N}
  P[M_{XY}]
  +\f{2\lm}{3}\f{N+1}{N}P[M_{X}]
  \label{eq:sdp}
\ee
Using these equations, the gap equation for the sigma condensate is given by
\bea
  \left. \f{d V(\ppc,G(\ppc))}{d\sigma_c}\right|_{\sigma_c=\sigma_0, \pi_c=0}
   &=& \left.\left(\f{\pa V(\ppc,G(\ppc))}{\pa \sigma_c}
  +\f{\pa V(\ppc,G(\ppc))}{\pa G}\f{d G(\ppc)}{d \sigma_c}\right)
  \right|_{\sigma_c=\sigma_0, \pi_c=0} \nn
  &=&\left.\f{\pa V(\ppc,G(\ppc))}{\pa \sigma_c}
  \right|_{\sigma_c=\sigma_0, \pi_c=0} \nn &=& 0
\eea
which gives
\be
  \sigma_0 \left[ m^2 + \f{2\lm}{3N}\sigma_0
  +\f{2\lm}{N}P[M_{XY}] 
  +\f{2\lm}{3}\f{N-1}{N}P[M_{X}] \right]
  =0
\ee
This equation reduces to the very simple form when the SD-equations
are used again.
\be    \label{eq:gap}
  \sigma_0^2 \left[ M_{XY}^2 - \f{4\lm}{3N}\sigma_0^2 
  \right] =0
\ee

Here we need some case for the meson masses.
The question is whether $M_\sigma$ and $M_\pi$, which are the parameters that
appear in the SD-equation, can be identified as the physical masses of 
the sigma and pions.
In fact, it is more appropriate to define the masses by the curvatures of the 
effective potential around its minimum, namely, 
\be   \label{eq:msigma-def}
  m^2_\sigma \equiv \left. \f{d^2 V}{d \sigma_c^2} \right|_{\sigma_c=\sigma_0,
  \pi_c =0}
\ee
\be   \label{eq:mpi-def}
  m^2_\pi \equiv \left. \f{d^2 V}{d \pi_c^2} \right|_{\sigma_c=\sigma_0,
  \pi_c =0}
\ee
rather than $M_\sigma$ and $M_\pi$.
The quantities $M_\sigma$ and $M_\pi$ are merely regarded as
variational parameters which are determined from the SD-equation.
In fact if higher order loop corrections are taken into account in 
$\Gamma_2$ in the CJT action,
the solutions of the SD-equation, $M_\sigma$ and $M_\pi$, become
momentum-dependent and then cannot be simply identified with the
physical meson masses.
Of course, the masses defined by eqs.(\ref{eq:msigma-def}) and 
(\ref{eq:mpi-def})
are also approximate values unless all the loop contributions
are taken into account.
But we see that the NG theorem is always satisfied because 
the $O(N)$ symmetry is satisfied at any order of expansion.
This way of defining the masses in the effective potential formalism 
is also seen in ref.\cite{Roh98,Oko96}
\footnote{In \cite{Oko96}, the dressed propagator ansatz
(\ref{eq:dress}) are used.
In this case even if one defines the meson masses by the second derivatives
of the effective potential, the NG theorem is violated in the
CJT formalism because of the violation of the $O(N)$ symmetry in the
effective potential.}.
The differences between two definitions come from the truncation of
the loop expansion of the CJT action.
Since the physical mass is defined by the pole of the two-point function,
if one can calculate the effective potential up to all-order,
both definitions might give the same mass.
Our definition of the mass is based on the approximation, which respects
the $O(N)$ symmetry and therefore the NG-theorem is satisfied at each
loop diagram of $\Gamma_2$ in the CJT action.

We evaluate the meson masses according to the definitions
(\ref{eq:msigma-def}) and (\ref{eq:mpi-def}).
The results are
\be     \label{eq:msigma-2}
  m_\sigma^2 = 2 \sigma_0^2 \left( \left.\f{dM_{XY}^2}{d(\ppc^2)}
  \right|_{\sigma_c=\sigma_0,\pi_c=0} - \f{4\lm}{3N} \right)
\ee
\be     \label{eq:mpi-2}
  m_\pi^2 = 0
\ee
in the broken symmetry phase.
The pion mass vanishes as expected.
The derivative term in the sigma meson mass is transformed into
\bea        \label{eq:deriv}
   \f{dM_{XY}^2}{d(\ppc^2)}
  &=& \left( 18\lm N + 8\lm^2 (N+2) P_2[M_{X}] \right) \nn
  & & \times
    \left( 9N^2+6\lm N(N+1)P_2[M_{X}]+18\lm NP_2[M_{XY}]
     \right. \nn
  & & \qquad \left.
      +8(N+2)\lm^2 P_2[M_{X}]P_2[M_{XY}] \right)^{-1}
\eea
with
\be         \label{eq:P2}
  P_2[M] \equiv \intbk \f{1}{(k^2+M^2)^2}
\ee
by the SD-equations.

Finally we comment on the differences between the conventional method and
ours again.
As we showed previously, the CJT effective potential has two arguments, i.e.,
$\ppc$ and $G$.
Then the chiral symmetry breaking may appear in two ways, one as a nonzero
$\p_c$ and the other as a nonsymmetric $G$.
The first one corresponds to the ordinary spontaneously symmetry breaking
and to our parametrization of $G$.
The latter case is that the vacuum expectation value of the two point function
breaks the chiral symmetry,
and the dressed propagator ansatz(\ref{eq:dress}) corresponds to this choice.
Since there occurs spontaneous symmetry breaking in both cases,
the NG bosons must appear.
In the latter case, however, they do not correspond to the pions,
as they are unphysically massive.
If one is interested in low-energy QCD phenomenology or SCSB, the former 
choice is the right way of the symmetry breaking
because the pions appear as the massless NG bosons, and the
ordinary effective potential is reproduced after eliminating $G$.
The difference between the two approaches might come
from the truncation of the loop
expansion of the effective potential, for both the approaches
should give the same physical mass in the full-loop calculation.

We discuss the treatment of these equations in a
phenomenological application in Sec. \ref{chp:app1} together with
numerical calculation.

\section{Renormalization}       \label{sec:renorm}

\subsection{A Conventional Cut-off Scheme}

  Renormalization of the CJT effective action in the linear sigma model
is discussed in ref.\cite{Ame97} in the past.
In that paper, divergent integrals are regularized with a cut-off and
the renormalized mass and coupling are defined in order to absorb the
divergent part.
This renormalization program was performed in the large $N$ limit\cite{CJP74},
where all the divergent terms are absorbed into the bare quantities.
However, if one applies this to the present case, one sees that
not all the divergences can be removed away.
We here review this renormalization following Amelino-Camelia
\cite{Ame97} and point out what is the problem.

We write the effective potential and the SD-equations for 
eq. (\ref{eq:dress}),
\bea
  V(\sigma_c,\pi_c) &=&
  \f{m^2}{2} (\sigma_c^2+\vec{\pi}_c^2)
  +\f{\lm}{6N}(\sigma_c^2+\vec{\pi}_c^2)^2 
  +\f{1}{2} Q[M_\sigma]+\f{N-1}{2}Q[M_\pi] \nn
  && -\f{1}{2}\left(M_\sigma^2-m^2-\f{2\lm}{N} \sigma_c^2
 -\f{2\lm}{3N}\vec{\pi}_c^2 \right)P[M_\sigma] \nn
 && -\f{1}{2}\left[ (N-1)\left(M_\pi^2-m^2-
 \f{2\lm}{3N}\sigma_c^2\right)-(N+1)\f{2\lm}{3N}\vec{\pi}_c^2 \right] 
  P[M_\pi] \nn
  && +\f{\lm}{2N}(P[M_\sigma])^2 + \f{\lm}{6}\f{N^2-1}{N}(P[M_\pi])^2
  + \f{\lm}{3}\f{N-1}{N} P[M_\sigma]P[M_\pi]
\eea
\be
  M_\sigma^2 = m^2 +\f{2\lm}{N} \sigma_c^2 
  +\f{2\lm}{N} P[M_\sigma] +\f{2\lm}{3}\f{N-1}{N}P[M_\pi]
\ee
\be
  M_\pi^2=m^2+\f{2\lm}{3N}\sigma_c^2
  +\f{2\lm}{3}\f{N+1}{N}P[M_\pi] +\f{2\lm}{3N}P[M_\sigma]
\ee
where we take the conventional $O(N)$ violated form of $G$ according to
\cite{Ame97}, but our discussion on the renormalization is unchanged 
if we take the $O(N)$ symmetric form of $G$.
These quantities are all written by bare quantities.
The loop integral $P[M]$, shown in eq.(\ref{eq:on-P}), of course diverges.
We regularize the temperature independent term with the cut-off $\Lambda$.
\be
  P[M] = I_1-M^2 I_2+P_f[M]
\ee
\be
  P_f[M]=\f{M^2}{16\pi^2}\ln\f{M^2}{\mu^2}-\int\f{d^3k}{(2\pi)^3}
  \left[\sqrt{\vec{k}^2+M^2}\left(1-\exp\left(\f{\sqrt{\vec{k}^2+M^2}}{T}
  \right)\right)\right]^{-1}
\ee
\be
  I_1 = \f{\Lambda^2}{8\pi^2},\quad
  I_2 = \f{1}{16\pi^2}\ln\f{\Lambda^2}{\mu^2}
\ee
where $\mu$ denotes the renormalization scale and is arbitrary.
Divergence also appears in the integral of logarithm in the effective 
potential.
\be
  \f{1}{2}Q[M] = -\f{M^4}{4}I_2+\f{M^2}{2}I_1+Q_f[M]
\ee
\be
  Q_f[M]=\f{M^4}{64\pi^2}\ln\f{M^2}{\mu^2}
  +T\ln \left[ 1-\exp\left(-\f{\sqrt{\vec{k}^2+M^2}}{T}\right)\right]
\ee

After the regularization, one must renormalize them.
Amelino-Camelia determines the renormalized mass $m_R$ and 
coupling $\lm_R$ as
\be
  \f{m_R^2}{\lm_R}=\f{m^2}{\lm}+\f{2}{3}\f{N+2}{N}I_1, \quad
  \f{1}{\lm_R}=\f{1}{\lm}+\f{2}{3}\f{N+2}{N}I_2
\ee
respectively.
This is, however, renormalized only at $M_\sigma=M_\pi$.
In fact when $M_\sigma=M_\pi\equiv M$ is hold,
we have
\be
  M^2 = m_R^2+\f{4}{3N}\lm_R \sigma_c^2 +\f{2\lm_R}{3}\f{N+2}{N}P_f[M]
\ee

However, if we use it in the broken phase, all the terms in the SD-equation and
the effective potential are
expressed not only by the renormalized quantities but bare quantities
remain in several terms.
Essentially the same situation is encountered in the mean field 
approach\cite{Roh98}.

Under these circumstances, Amelino-Camelia makes the following arguments.
If we are interested in the possibility of renormalization,
we drop the $\lm$ term in the infinite limit of the cut-off.
We then get, however, $M_\sigma=M_\pi$.
He also says that if one treats the model as the low-energy effective model,
the cut-off $\Lambda$ should be a large finite value which is smaller than the
Landau pole.
He concludes that $M_\pi$ vanishes at $\lm\to 0$ but remains finite
for the finite value of the cut-off.


We agree that the $O(N)$ linear sigma model in four dimensions
may be trivial if the full quantum effects are included. 
Our analysis is however basically one-loop
level and therefore semi-classical. So we consider another way of the
renormalization which removes divergences order by order both in the
symmetry restored phase and broken phase.

  Recently a regularization scheme, which is similar to the
Pauli-Villars regularization, is used in the CJT action \cite{Len99}.
We here give a new renormalization scheme based on auxiliary field
method with the dimensional regularization.

\subsection{An Application of the Auxiliary Field Method}

  In order to solve the above problem of renormalization of the CJT
effective action,
we here use the renormalization procedure of the 
auxiliary field.
In the original CJT's paper\cite{CJT74}, 
it is reported that the effective potential
in the large $N$ limit in the CJT formalism is the same form obtained from
the auxiliary field method\cite{CJP74}.
Since the solutions of the SD-equation, $X$ and $Y$, are momentum 
independent here, we can regard them as an auxiliary field.
Then one can renormalize the CJT effective
potential with the auxiliary field method following some 
literature\cite{CJP74,Kob75}.

Following the above argument, we first rewrite the $O(N)$ symmetric
CJT effective 
potential(\ref{eq:CJTeff-2}) using the SD-equations(\ref{eq:sds}) and
(\ref{eq:sdp})
\bea
  \lqn{V(\ppc,M_{XY},M_{X}) } \nn
  &=& \f{1}{2}M_{XY}^2 \ppc^2 - \f{\lm}{3N} \ppc^4
  + \f{3N}{4\lm (N+2)}M_{XY}^2 m^2 
  + \f{3N (N-1)}{4\lm(N+2)}M_{X}^2 m^2 \nn
  & & -\f{3N}{16\lm (N+2)} \left\{ (N+1)M_{XY}^4
      +3(N-1)M_{X}^4 \right .\nn
  & & \qquad\qquad\qquad\qquad \left.-2(N-1)M_{XY}^2 M_{X}^2
      +2Nm^4 \right\} \nn
  & & +\f{1}{2}\intk \ln(k^2+M_{XY}^2)
      +\f{N-1}{2}\intk \ln(k^2+M_{X}^2)
\eea
Throughout this section, we ignore the temperature dependent terms
since they are all finite and not renormalized.
This means that the loop integrals in the equations are simply four-momentum
integrals.
We consider that the extension to the system at finite temperature
should be carried out for the well-defined effective potential which
has no divergence after renormalization.
Once the renormalization is carried out at $T=0$,
it is straightforward to extend it to the system at finite temperature
since the temperature dependent terms all converge in the UV limit due
to the Boltzmann factor.

We add four counter terms in order to cancel the divergence,
\be
  A M_{XY}^2 m^2 + B (N-1) M_{X}^2 m^2
  +\f{1}{2} C M_{XY}^4 + \f{N-1}{2}D M_{X}^4
\ee
where $A, B, C$ and $D$ are renormalization constants to adjust the
divergences and become
\be
  A=B=0, \qquad C=D=\f{1}{32\pi^2}\left(\f{2}{4-d}-\gamma_E+
  \ln (4\pi) \right)
\ee
if the $\overline{MS}$ scheme of dimensional regularization is used.
These correspond to the following renormalization conditions.
\be
  \left. \f{dV}{d M_{XY}^2} 
  \right|_{\sigma_c^2=0,M_{XY}^2=M_{X}^2=0}
  = \f{3Nm^2}{4\lm(N+2)}
\ee
\be
  \left. \f{dV}{d M_{X}^2} 
  \right|_{\sigma_c^2=0,M_{XY}^2=M_{X}^2=0}
  = \f{3N(N-1)m^2}{4\lm(N+2)}
  \label{eq:rencon1}
\ee
\be
  \left. \f{d^2V}{(d M_{XY}^2)^2} 
  \right|_{\sigma_c^2=0,M_{XY}^2=M_{X}^2=\mu^2}
  = -\f{3N(N+1)}{8\lm(N+2)}
\ee
\be
  \left. \f{d^2V}{(d M_{X}^2)^2} 
  \right|_{\sigma_c^2=0,M_{XY}^2=M_{X}^2=\mu^2}
  = -\f{9N(N-1)}{8\lm(N+2)}
  \label{eq:rencon2}
\ee
respectively.
Therefore the renormalized effective potential is given by
\bea
  \lqn{V(\ppc,M_{XY},M_{X}) } \nn
  &=& \f{1}{2}M_{XY}^2 \ppc^2 - \f{\lm}{3N} \ppc^4
  + \f{3N}{4\lm (N+2)}M_{XY}^2 m^2 
  + \f{3N (N-1)}{4\lm(N+2)}M_{X}^2 m^2 \nn
  & & -\f{3N}{16\lm (N+2)} \left\{ (N+1)M_{XY}^4
      +3(N-1)M_{X}^4 \right .\nn
  & & \qquad\qquad\qquad\qquad \left.-2(N-1)M_{XY}^2 M_{X}^2
      +2Nm^4 \right\} \nn
  & & +\f{1}{64\pi^2}\left(\ln\f{M_{XY}^2 }{\mu^2}-\f{3}{2}\right)
      +\f{N-1}{64\pi^2}\left(\ln\f{M_{X}^2 }{\mu^2}-\f{3}{2}\right)
  \label{eq:ren-vren}
\eea
The renormalized SD-equations are the same form as eqs.(\ref{eq:sds}),
(\ref{eq:sdp}) except for the divergent terms.
\be
  M_{XY}^2
  = m^2+\f{2\lm}{N}\ppc^2+\f{2\lm}{N}\f{M_{XY}^2}{16\pi^2}
  \left( \ln\f{M_{XY}^2}{\mu^2}-1\right) 
  +\f{2\lm}{3}\f{N-1}{N}\f{M_{X}^2}{16\pi^2}
  \left( \ln\f{M_{X}^2}{\mu^2}-1\right)
  \label{eq:sds-r}
\ee
\be
  M_{X}^2
  = m^2+\f{2\lm}{3N}\ppc^2+\f{2\lm}{3N}\f{M_{XY}^2}{16\pi^2}
  \left( \ln\f{M_{XY}^2}{\mu^2}-1\right) \nn
  +\f{2\lm}{3}\f{N+1}{N}\f{M_{X}^2}{16\pi^2}
  \left( \ln\f{M_{X}^2}{\mu^2}-1\right)
  \label{eq:sdp-r}
\ee
The gap equation for the sigma condensate and the formulae of the meson masses
are the same form as eqs.(\ref{eq:gap}), (\ref{eq:msigma-2}) and
(\ref{eq:mpi-2}).
They are all finite because $M_{XY}$ and $M_{X}$ are finite.

  Here we make a short remark on the problem in the large $N$ limit.
In refs.\cite{Kob75,Abb76} the stability of the effective potential of the 
linear sigma model in the large $N$ limit was discussed.
They found that the true minimum of the effective potential gives
the trivial phase in the large $N$ limit.
Although the renormalization scheme they use is different from ours,
there exists the same situation in the present approach.
In fact one can show that their results are identical to ours
in the large $N$ limit.
Also in this limit, $M_{XY}$ vanishes and the
renormalization conditions (\ref{eq:rencon1}) and (\ref{eq:rencon2})
correspond to eqs.(2.5) and (2.6) in ref.\cite{Kob75}, respectively.

\subsection{Summary}

  We summarize the results of this section.
First we formulate the $O(N)$ linear sigma model with the CJT effective
action.
The Hartree-Fock approximation is used, which corresponds to incorporating
only the double bubble diagram in the two-loop level in the CJT formalism.
We adopt the $O(N)$ symmetric form of the propagator $G$, otherwise the
NG theorem is violated.
On the renormalization, we use the method of the auxiliary field.
One can remove all the divergence in the effective potential even in the
broken phase in this method.
The effective potential is then eq.(\ref{eq:ren-vren}) plus the thermal
effects.
\bea
  \lqn{V(\ppc,M_{XY},M_{X}) } \nn
  &=& \f{1}{2}M_{XY}^2 \ppc^2 - \f{\lm}{3N} \ppc^4
  + \f{3N}{4\lm (N+2)}M_{XY}^2 m^2 
  + \f{3N (N-1)}{4\lm(N+2)}M_{X}^2 m^2 \nn
  & & -\f{3N}{16\lm (N+2)} \left\{ (N+1)M_{XY}^4
      +3(N-1)M_{X}^4 \right .\nn
  & & \qquad\qquad\qquad\qquad \left.-2(N-1)M_{XY}^2 M_{X}^2
      +2Nm^4 \right\} \nn
  & & +\f{1}{2}Q_T[M_{XY}^2]
      +\f{N-1}{2}Q_T[M_{X}^2] \nn
  & & +\f{1}{64\pi^2}\left(\ln\f{M_{XY}^2}{\mu^2}-\f{3}{2}\right)
      +\f{N-1}{64\pi^2}\left(\ln\f{M_{X}^2}{\mu^2}-\f{3}{2}\right)
  \label{eq:ren-vfull}
\eea
with
\be
  Q_T[m] = \int \f{d^3k}{(2\pi)^3} 
  \f{2}{\beta}\ln(1-e^{-\beta\omega})
\ee
$M_{X}$ and $M_{XY}$ are determined by the SD-equations
\bea
  M_{XY}^2
  &=& m^2+\f{2\lm}{N}\ppc^2 + \f{2\lm}{N} P_T[M_{XY}]
      +\f{2\lm}{3}\f{N-1}{N} P_T[M_{X}] \nn
  & & +\f{2\lm}{N}\f{M_{XY}^2}{16\pi^2}
  \left( \ln\f{M_{XY}^2}{\mu^2}-1\right) 
  +\f{2\lm}{3}\f{N-1}{N}\f{M_{X}^2}{16\pi^2}
  \left( \ln\f{M_{X}^2}{\mu^2}-1\right)
  \label{eq:ren-sdmxy}
\eea
\bea
  M_{X}^2
  &=& m^2+\f{2\lm}{3N}\ppc^2+\f{2\lm}{3N}P_T[M_{XY}]
      +\f{2\lm}{3}\f{N+1}{N} P_T[M_{X}] \nn
  & & +\f{2\lm}{3N}\f{M_{XY}^2}{16\pi^2}
  \left( \ln\f{M_{XY}^2}{\mu^2}-1\right) 
  +\f{2\lm}{3}\f{N+1}{N}\f{M_{X}^2}{16\pi^2}
  \left( \ln\f{M_{X}^2}{\mu^2}-1\right)
  \label{eq:ren-sdmx}
\eea
with
\be
  P_T[m] = \int \f{d^3k}{(2\pi)^3}\f{n_B(\omega_k)}{\omega_k}
\ee
The sigma condensate is given by eq.(\ref{eq:gap}) which is
finite since $M_{XY}$ is finite in this case.

  The meson masses are also given by eqs.(\ref{eq:msigma-2})
and (\ref{eq:mpi-2}).
The pion mass is of course massless in the broken symmetry phase.
In the equation of the sigma meson mass, $P_2[m]$ factor is replaced
by
\be
  P_2[m] \to P_{2,T}[m] - \f{1}{16\pi^2}\ln\f{m^2}{\mu^2}
\ee
\be
  P_{2,T}[m] = \f{\beta}{2}\int\f{d^3k}{(2\pi)^3}\f{1}{\omega_k^2}
 \left[\left(1+\f{1}{\beta\omega_k}\right)
 n_B(\omega_k)+n_B(\omega_k)^2\right]
  \label{eq:high-p2tf}
\ee
in the $\overline{MS}$ scheme of the dimensional regularization.

\section{An Application to the Low-energy Mesons}
   \label{chp:app1}

In this and the following sections, we apply the CJT formalism introduced
in Sec. \ref{chp:on} to systems of low-energy hadrons.
We include both the thermal and quantum corrections in numerical 
calculation.

\subsection{Formulation}

  First we formulate a system with a sigma meson and pions.
Since the real pions have masses because of the small but non-zero
$u, d$ quark masses,
we introduce a term which explicitly breaks chiral symmetry in the 
lagrangian.
\be
  \cl = \f{1}{2}\pa_\mu \p^a \pa_\mu \p^a +\f{1}{2} m^2 \p^2
        +\f{\lm}{6N}(\p^2)^2 - c \sigma
\ee
with $\p^1=\sigma$.
Since the two-point function is obtained from the second-derivative of the 
lagrangian, this term does not affect it and 
the SD-equations for $M_{XY}$ and $M_{X}$ are the same forms as
eqs.(\ref{eq:ren-sdmxy}) and (\ref{eq:ren-sdmx}).
Thus the effective potential is the same as eq.(\ref{eq:ren-vfull}) except for
the $c \sigma$ term.
\bea
  V(\ppc,M_{XY},M_{X})
  &=& \f{1}{2}M_{XY}^2 \ppc^2 - \f{\lm}{3N} \ppc^4-c\sigma \nn
  & & + \f{3N}{4\lm (N+2)}M_{XY}^2 m^2 
      + \f{3N (N-1)}{4\lm(N+2)}M_{X}^2 m^2 \nn
  & & -\f{3N}{16\lm (N+2)} \left\{ (N+1)M_{XY}^4
      +3(N-1)M_{X}^4 \right .\nn
  & & \qquad\qquad\qquad\qquad \left.-2(N-1)M_{XY}^2 M_{X}^2
      +2Nm^4 \right\} \nn
  & & +\f{1}{2}Q_T[M_{XY}^2]
      +\f{N-1}{2}Q_T[M_{X}^2] \nn
  & & +\f{1}{64\pi^2}\left(\ln\f{M_{XY}^2}{\mu^2}-\f{3}{2}\right)
      +\f{N-1}{64\pi^2}\left(\ln\f{M_{X}^2}{\mu^2}-\f{3}{2}\right)
  \label{eq:ren-vexp}
\eea

The gap equation for the sigma condensate in turn becomes
\be    
  \sigma_0^2 \left[ M_{XY}^2(\sigma_0^2) - \f{4\lm}{3N}\sigma_0^2 
  \right] =c
  \label{eq:app1-gap-exp}
\ee
From this equation one sees that the pion mass does not vanish even below the
critical point
\be     
  m_\pi^2 = \f{c}{\sigma_0}
  \label{eq:app1-mpi-exp}
\ee
and the sigma meson mass is given by
\be     
  m_\sigma^2 = 2 \sigma_0^2 \left( \left.\f{dM_{XY}^2}{d(\ppc^2)}
  \right|_{\sigma_c=\sigma_0,\pi_c=0} - \f{4\lm}{3N} \right) + 
  \f{c}{\sigma_0}
  \label{eq:app1-msigma-exp}
\ee
where the derivative term is the same form as eq.(\ref{eq:deriv}).

\subsection{Numerical Results}

  Here we show some numerical results of the above equations.
We first concentrate only on the thermal effects and ignore the
quantum corrections.
This approximation is frequently used in literature 
when one is interested in the thermal effects.
We will include the quantum corrections in the next subsection.
Throughout this and the next section, we set $N=4$, i.e.,
the $SU(2)_L\times SU(2)_R$ chiral symmetry which corresponds to the real world
of one sigma and three pions.

\subsubsection{Thermal Corrections}
  \label{subsc:therm1}

We use the values at zero temperature as initial conditions of numerical
calculation.
Because we ignore the quantum corrections at this stage, the solutions of the SD-equations $M_{XY}$ and $M_X$ are 
equal to $m_\sigma$ and $m_\pi$ at $T=0$ as in the chiral limit.
Initial parameters are determined from the condition at $T=0$:
\be
  c = \fpi m_\pi^2(T=0)
  \label{eq:gmor}
\ee
\be
  \lm = \f{3}{\fpi^2} (m_\sigma^2(T=0) - m_\pi^2(T=0) )
\ee
\be
  m^2 = -\f{1}{2}m_\sigma^2(T=0) + \f{3}{2}m_\pi^2(T=0)
\ee
where $f_\pi(=93$ MeV) is the pion decay constant at zero temperature.

For the mass of $\sigma$,
Particle Data Group gives the value of
400-1200 MeV\cite{PDG}.
We take $m_\sigma=600$ MeV as typical values.
Qualitative properties change little if the mass of the sigma meson
is to be $m_\sigma=1$ GeV.
$m^2$ is negative because the spontaneous symmetry breaking takes place.

 These results are essentially the same as ref.\cite{Pet98} because we
ignore the quantum corrections here.
In ref.\cite{Pet98}, however, the author identifies $M_{XY}$ and $M_X$ with
the sigma meson mass and the pion mass, respectively, so that he 
concludes that the NG theorem is violated at finite temperature.
As discussed before, we do not regard these values as the physical
meson masses and rather consider them simply as variational parameters.
We calculate the sigma meson mass according to eq.(\ref{eq:app1-msigma-exp})
instead.

  When we evaluate the sigma meson mass in the chiral limit in this
approximation, however, we encounter a difficulty due to  the infrared
singularity.
As shown in Subsec. \ref{subsec:ours}, there is a factor $P_2[m]$ in the
equation of the sigma meson mass eq.(\ref{eq:app1-msigma-exp}).
Its temperature-dependent part $P_{2,T}[m]$ has the form
\bea
  \lqn{P_{2,T}[m]} \nn
  &=& \f{1}{4\pi^2}\int_0^\infty dx \f{x^2}{x^2+\left(\f{m}{T}\right)^2}
  \left[\left(1+\f{1}{\sqrt{x^2+\left(\f{m}{T}\right)^2}}\right)
  \f{1}{e^{\sqrt{x^2+\left(\f{m}{T}\right)^2}}-1}
  +\f{1}{\left(e^{\sqrt{x^2+\left(\f{m}{T}\right)^2}}-1\right)^2}\right] \nn
  &&
\eea
We naively suppose that this temperature dependent part vanishes at
$T\to 0$.
This is indeed so for $m\neq 0$.
However, this factor gives a non-zero contribution
if $m$ approaches to zero for $T\to 0$ keeping $m/T$ finite.
This phenomenon actually occurs in the strict chiral limit if we neglect
quantum corrections,
because one of the solutions of the SD-equation, $M_{X}$, vanishes
at $T=0$.
The fact that $M_{X}/T$ approaches to a non-zero value 
at $T=0$ can be confirmed by numerical calculation.
Thus the sigma meson mass $m_\sigma (T\to 0)$ 
deviates from $m_\sigma (T= 0)=600$ MeV.
We consider that this is because we have taken only the temperature
dependent part in $P_2[m]$.
In fact, in the chiral limit,
the temperature-independent part 
$P_{2,Q}[m]$ is also infrared divergent at $T=0$ and is to cancel
the infrared singularity in $P_{2,T}[m]$.
So it may not be possible to separate the temperature-dependent part and
the temperature-independent one in $P_2[m]$.

  There are two possibilities to solve this problem.
One is to introduce the explicit chiral symmetry breaking term $-c\sigma$,
as seen above.
This makes $M_{X}$ finite at $T=0$, corresponding
to the finite-mass pions.
So if we take the pion mass to be considerably large, e.g., 
$m_\pi(T=0)=138$ MeV, this problem apparently disappears.
Another possibility, which we think is better, is to include quantum 
corrections, which makes $M_{X}$ finite even in the chiral limit.
The study along this line is given in the following subsection.

\subsubsection{Thermal and Quantum Corrections}
  \label{subsub:qc}

  In this subsection, we include the quantum corrections following the
renormalization technique of the auxiliary field discussed in Sec.
\ref{sec:renorm}.
We have then another parameter $\mu$ which is introduced in the dimensional
regularization and corresponds to the renormalization scale.
As $\mu$ is a free parameter,
we must choose a suitable value $\mu$ at which the spontaneous
symmetry breaking actually occurs.
All the other parameters are determined at the chosen $\mu$
as reasonable values.
Before proceeding to finite temperature, we perform a simple estimation
in our model.

In QCD, the following relation is well satisfied.
\be
  \fpi^2 m_\pi^2 = 2 m_q \la \overline{q} q \ra
\ee
where $q$ denotes the light $u, d$ quarks.
This relation is called the Gell-Mann--Oakes-Renner(GMOR) relation and is
derived from the low-energy theorem of QCD.
In the chiral perturbation theory  which is one of the low-energy effective
models of QCD, this relation is also satisfied to the first order in the 
quark mass.
This equation shows that the pion mass square is proportional to the
quark mass.
The deviation from it corresponds to higher order effects of the chiral
perturbation theory.
Since the chiral perturbation theory is a perturbation expansion in both
external momenta of the NG bosons and the current quark masses,
if this relation is satisfied up to the relatively large quark mass,
it means that the chiral perturbation theory is valid up to this momentum 
scale.
We recently investigated the behavior of the GMOR relation by
constructing the pion as a relativistic bound state of quark and antiquark
pair using the Bethe-Salpeter equation\cite{Nai98b}.
We showed that
the GMOR relation is quite well satisfied at least up to the strange quark mass
($\sim 150$ MeV) region.

\begin{figure}[h]
  \hspace*{3cm}
  \epsfxsize=300pt
  \epsfbox{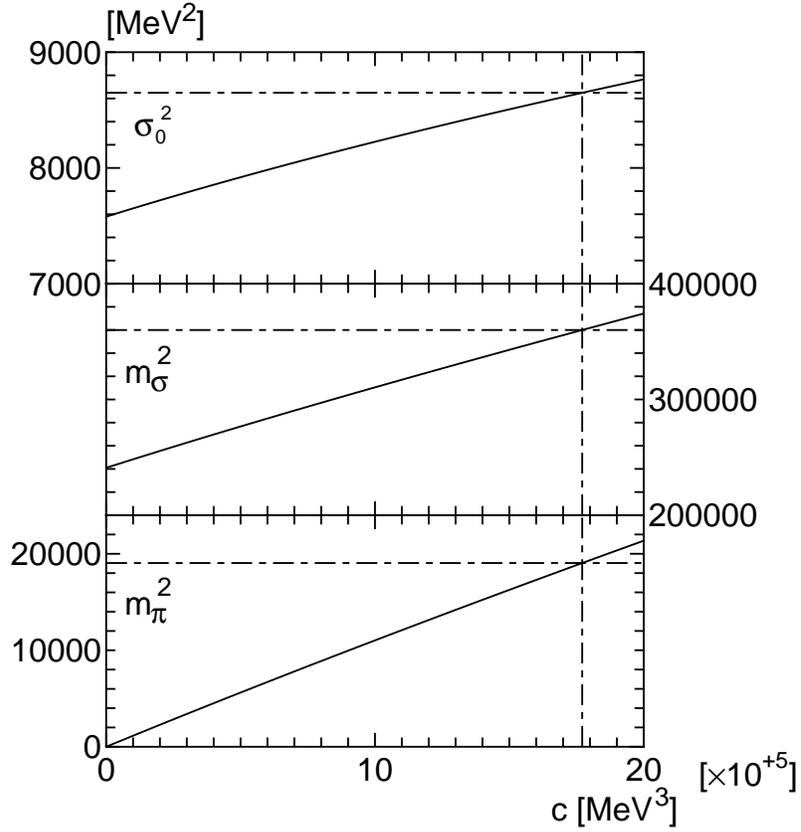}
  \caption{$\sigma_0^2$, $m_\sigma^2$ and $m_\pi^2$ for $\mu=320$ MeV}
  \label{fig:set}
\end{figure}

In our linear sigma model the relation corresponding to GMOR relation is
given by eq.(\ref{eq:gmor}).
We can evaluate the $c$-dependence of $m_\pi^2$ in the following way.
First we determine the values of $c$ and the 
other parameters at $m_\pi=138$ MeV.
Then by changing the value of $c$ with the other parameters fixed,
we calculate $M_{XY}$ and $M_X$, and the meson masses.
The obtained results are shown in Fig. \ref{fig:set}.
We have confirmed that these results are almost independent of the 
renormalization scale.
These three values show almost linear dependence on $c$.
In the chiral limit ($c=0$),
the sigma condensate which corresponds to the pion decay constant
is about 87 MeV.
This is very similar to the one obtained in the chiral perturbation
theory\cite{Ga84} which gives 88 MeV.
The sigma meson is about 490 MeV
(we have fixed $m_\sigma = 600$ MeV at $m_\pi=138$ MeV) and
the pion mass of course vanishes in the chiral limit.
Thus we see that the linear sigma model gives consistent results
with the chiral perturbation theory.

Now we proceed to finite temperature.
Once the quantum corrections are included, $M_{XY}$ and $M_X$ do not
agree with the sigma and pion masses even at $T=0$.
We need to determine them by solving the SD-equations and the gap
equation for the sigma condensate.
We show the results of the initial parameters obtained in this way in
Table. \ref{tab:ini1}-\ref{tab:ini3}.
The value $T_c'$ is the point where the $\sigma_0$ curve crosses the 
temperature axis.
We see that for smaller $\mu$, $M_X$ can be negative so that the effective
potential becomes complex unphysical.
For larger $\mu$, $m^2$ becomes
positive so that the spontaneous symmetry breaking does not occur.
As a result a proper range of $\mu$ is naturally determined as is given in the
table.
Because the parameter $\mu$ is free within this range,
we show as an example a result for $\mu=320$ MeV in the case
of $m_\sigma(T=0)=600$ MeV.

\begin{figure}[t]
  \hspace*{3cm}
  \epsfxsize=300pt
  \epsfbox{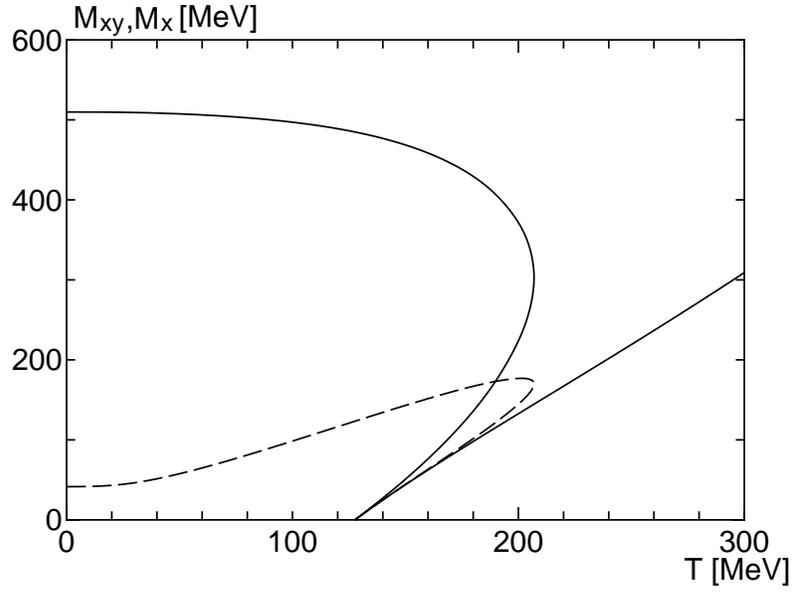}
  \caption{The solutions of SD-eqs. for $m_\sigma(T=0)=600$ MeV and
  $\mu=320$ MeV.
  The solid line is $M_{XY}$ and the dashed one is $M_X$}
  \label{fig:sd320600}
\end{figure}

\begin{figure}[h]
  \hspace*{3cm}
  \epsfxsize=300pt
  \epsfbox{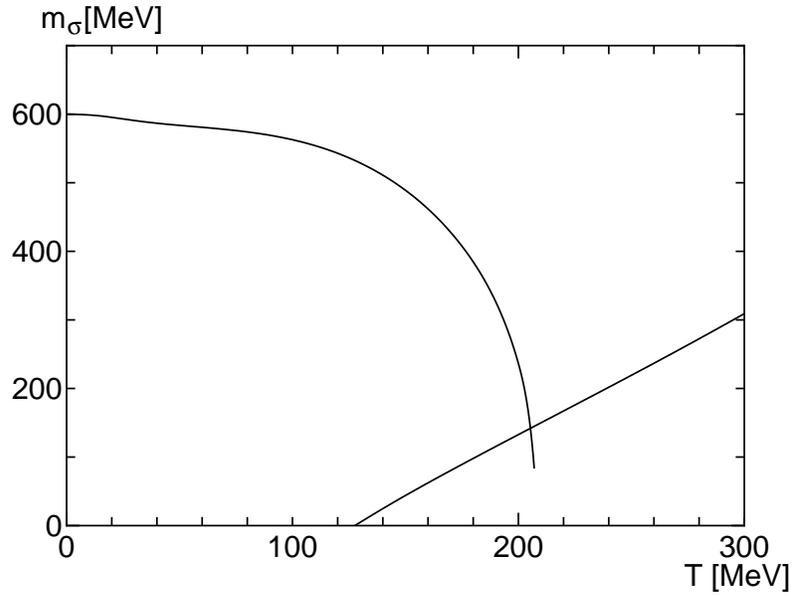}
  \caption{$m_\sigma$ for $m_\sigma(T=0)=600$ MeV and
  $\mu=320$ MeV}
  \label{fig:msigma320600}
\end{figure}

\begin{figure}[t]
  \hspace*{3cm}
  \epsfxsize=300pt
  \epsfbox{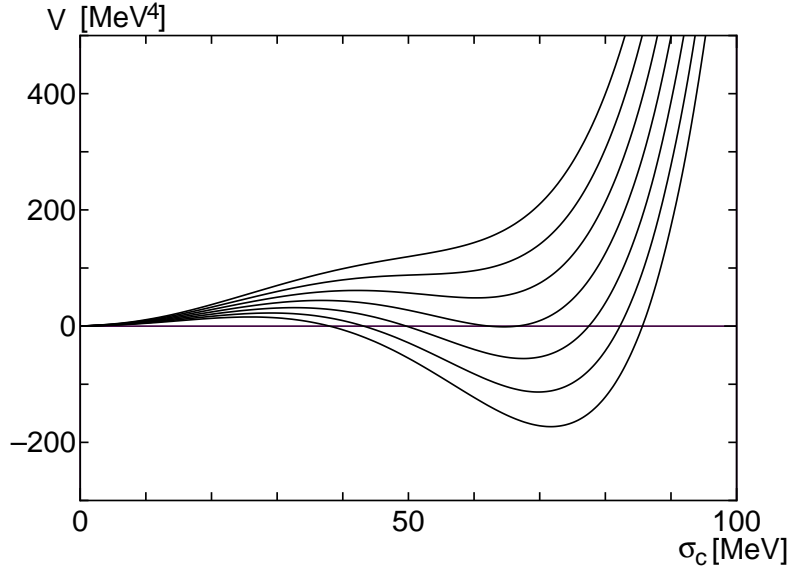}
  \caption{Effective potential for
  $m_\sigma(T=0)=600$ MeV and $\mu=320$ MeV.
   Solid lines denote the case of
  $T=$198 MeV, 194 MeV, $\cdots$,174 MeV from top to bottom.}
  \label{fig:pot320600}
\end{figure}

\begin{figure}[t]
  \hspace*{3cm}
  \epsfxsize=300pt
  \epsfbox{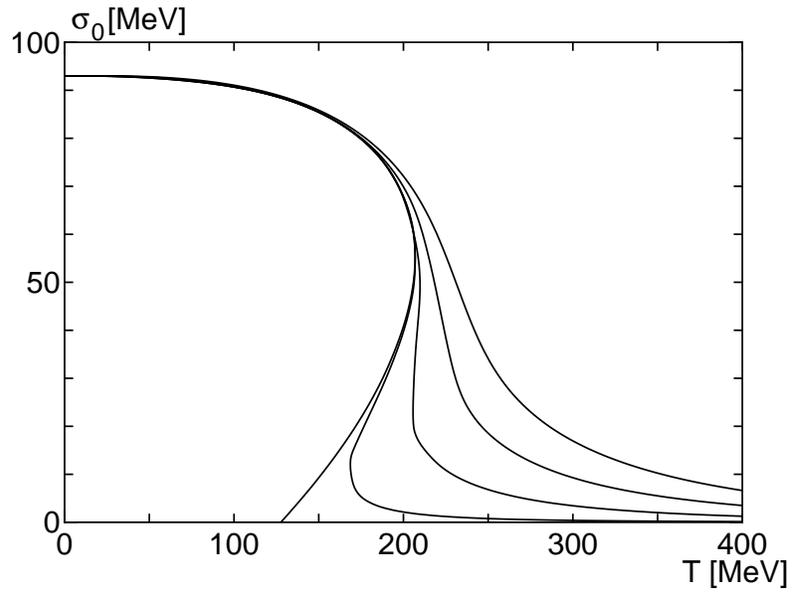}
  \caption{Sigma condensates $\sigma_0$ for $m_\sigma(T=0)=600$ MeV,
  $\mu=320$ MeV and $m_\pi(T=0)=0,20,60,100,138$ MeV from left to right.}
  \label{fig:con320set}
\end{figure}

We show the temperature dependence of the solutions of the SD-equations,
the physical meson masses and the effective potential in the chiral limit
in Figs. \ref{fig:sd320600}-\ref{fig:pot320600}.
Temperature dependence of the sigma condensate is also shown in Fig.
\ref{fig:con320set}.
The effective potential shows the signature of the first order phase
transition, whose behavior is similar to the case without quantum corrections.

We see also from the solutions of the SD-equations and the sigma 
condensate that the equations have two solutions at some temperature.
This is a typical feature of the first order phase transition, that is,
the upper solution corresponds to the (local) minimum of the potential and the 
lower to the maximum of the potential.

As the temperature increases from zero, the sigma condensate decreases, 
jumps to zero at $T_c$ and remains zero above $T_c$.
Likewise, in the figures $M_{XY}$ ($M_X$) changes along the solid 
(dashed) line, jumps to the lower solid line at $T_c$ and increases along
this line.

  If the deconfinement transition takes place at the same critical temperature
as the chiral phase transition, there are free quarks and gluons for $T > T_c$.
But recent lattice simulations suggest that there are some correlations of
quarks and gluons above $T_c$ \cite{det87}.
An instanton liquid model also favors this result \cite{shy}.
This may imply there exist hadronic excitations even above $T_c$.

The fact that the phase transition is the first order agrees with
other mean field approaches\cite{Bay77,Lar86,Roh98}.
It is, however, generally believed mainly from the analysis of the 
renormalization group equation
that the actual phase transition of the
$O(4)$ linear sigma model should
be of the second order.
We comment on this problem in the following section.

The critical temperature is about 184 MeV for $\mu=320$ MeV and about
178 MeV for $\mu=500$ MeV.
For the case of $m_\sigma(T=0)=1$ GeV, it is about 220 MeV for 
$\mu=400$ MeV and about 230 MeV for $\mu=580$ MeV.
So the renormalization scale affects the critical temperature little.
These values are reasonable since other approaches in the linear sigma
model predict similar values.

  We note that there is no peculiar behavior in the curves of the sigma meson 
mass, shown in Fig. \ref{fig:msigma320600}.
This is because $M_X$ does not vanish even at $T=0$ due to the quantum
corrections, shown in Fig. \ref{fig:sd320600}.
On the other hand, the pions remain massless below the critical 
temperature even in this case.

\begin{figure}[t]
  \hspace*{3cm}
  \epsfxsize=300pt
  \epsfbox{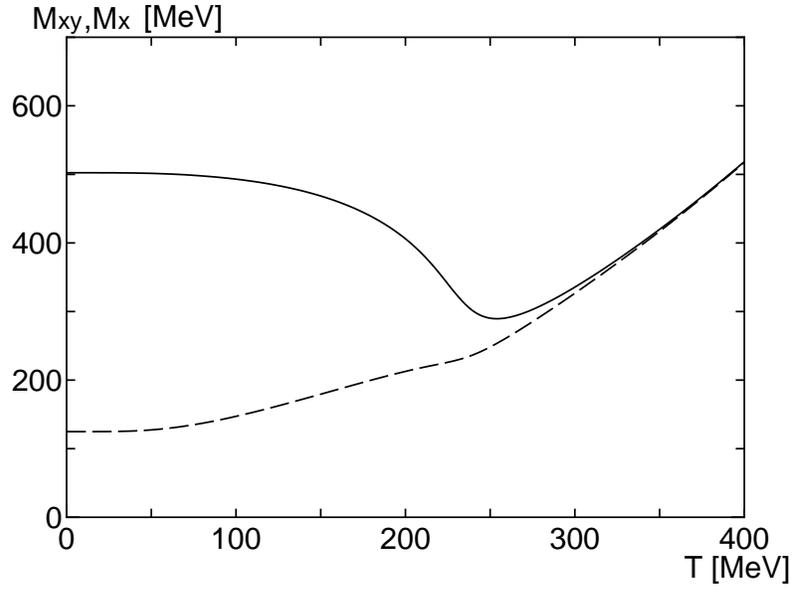}
  \caption{The SD-solutions $m_\sigma(T=0)=600$ MeV,
  $m_\pi(T=0)=138$ MeV and $\mu=320$ MeV.
  The solid line is $M_{XY}$ and the dashed one is $M_X$}
  \label{fig:sd320138}
\end{figure}

\begin{figure}[h]
  \hspace*{3cm}
  \epsfxsize=300pt
  \epsfbox{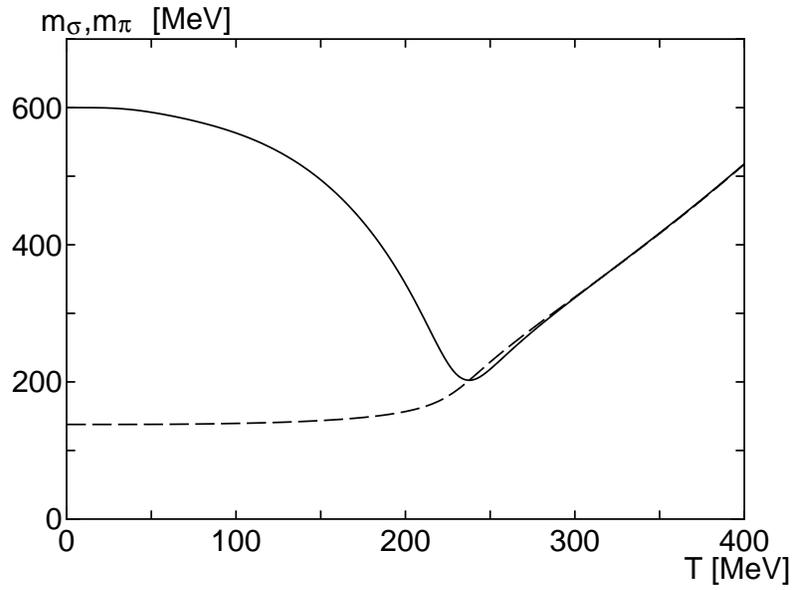}
  \caption{$m_\sigma$ for $m_\sigma(T=0)=600$ MeV, $m_\pi(T=0)=138$ MeV and
  $\mu=320$ MeV}
  \label{fig:m320138}
\end{figure}

Finally we show the cases with the explicitly chiral symmetry breaking,
i.e. $m_\pi(T=0)=138$ MeV in Figs. \ref{fig:sd320138}-\ref{fig:con320set}

\section{Comments on Effects of Other Loops}
   \label{sec:comm}

  We have so far calculated the physical quantities in the Hartree-Fock
approximation.
This means that the solutions of the SD-equations for the propagator,
$M_{XY}$ and $M_X$, are momentum-independent.
In the super-daisy diagrams, this assumption is quite natural because
no external momentum enters into loops.
In fact many authors use this approximation within the formalism of the
super-daisy diagrams in the mean field approach\cite{Bay77,Lar86,Roh98}.

  However, note that near the critical temperature there is no reason
to believe that only the super-daisy diagrams are dominant.
Within the framework of the Hartree-Fock approximation,
the first-order phase transition takes place as in our case.
(If we take the large $N$ limit in the CJT approach, 
the result is the second-order, see ref.\cite{Pet98}).
But the analyses of the renormalization group equations etc.
suggest that the phase transition of the
$O(4)$ linear sigma model is the second-order\cite{Raj93,Ogu98}.
Arnold and Espinosa pointed out that other loop diagrams than the
super-daisy diagrams are important near the critical temperature
\cite{Arn93}.
So the result for the order of the transition cannot be trusted in the
calculation with the super-daisy diagrams only.

\begin{figure}[h]
  \hspace*{5cm}
  \epsfxsize=100pt
  \epsfbox{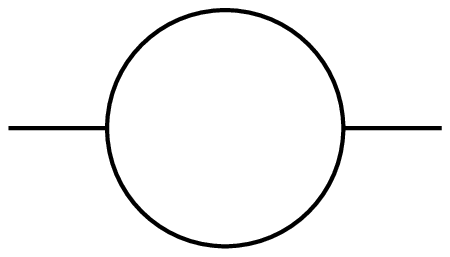}
  \caption{The loop diagram contributing to the two-point Green function
  which comes from the setting-sun type diagram.}
  \label{fig:ssun}
\end{figure}

  The next leading loop diagram contributing to the effective potential
is a setting-sun(SS) type diagram.
This diagram contributes to the two-point Green function as in Fig.
\ref{fig:ssun}.

It is obvious that the external momentum enters into the loop and therefore
the self-energy depends on the external momentum.
In our CJT formalism, this means that $M_{XY}$ and $M_X$ become
momentum-dependent.
Furthermore, we may have to take into account the wave function
renormalization by this diagram.
As far as we know, there is no calculation in which this diagram is
incorporated consistently.
Here we briefly estimate the contribution of the SS diagram
with an approximation in order to see whether the order of the transition
changes or not.
The term of the effective potential 
corresponding to the SS diagram is given by
\be
  -\f{4\lm^2}{9N^2}\ppc^a\ppc^c
  {\textstyle \sum} \hspace{-1em} \int_{\beta,p}
  {\textstyle \sum} \hspace{-1em} \int_{\beta,q}\left\{
  2 G_{ad}(p)G_{bd}(q)G_{bc}(-p-q)+G_{ac}(p)G_{bd}(q)G_{bd}(-p-q)\right\}
\ee
Hereafter we consider a simple case with $N=1$ since we are now interested in 
the phase transition rather than the hadron phenomenology.
The CJT effective potential is now given by
\bea
  V(\ppc,G) &=& \f{1}{2}m^2\ppc^2 + \f{\lm}{24}\ppc^4 + 
  \f{1}{2} \intbk \ln G^{-1}(k) +\f{1}{2}\intbk \cd^{-1}(k)G(k) \nn
  & & +\f{\lm}{8}\left(\intbk G(k)\right)^2 
      -\f{\lm^2}{12}\ppc^2
  {\textstyle \sum} \hspace{-1em} \int_{\beta,p}
  {\textstyle \sum} \hspace{-1em} \int_{\beta,q}G(p)G(q)G(-p-q)
\eea
with
\be
  \cd^{-1}(k) = k^2 + m^2 + \f{\lm}{2}\ppc^2
\ee
From the extremal condition, we obtain the SD-equation for $G$.
\be
  G^{-1}(p) = \cd^{-1}(p) + \f{\lm}{2}\intbk G(k) -
  \f{\lm^2}{2}\ppc^2\intbk G(k)G(-p-k)
  \label{eq:sdset}
\ee
Since the integral of the last term in eq.(\ref{eq:sdset}) depends on 
the external momentum, it is natural to put the form of $G$ as
\be
  G(p) = \f{A(p)}{p^2+M(p)^2}
\ee
Here we take a very simple form in which any external momentum dependence
in the parameters are neglected, i.e., $A=1$ and $M(p)=M$.
It can be calculated relatively easily in the CJT formalism by this procedure.
Though
this is rather a strong approximation and seems to be similar to 
the super-daisy approximation,
we see that the order of the phase transition can change in some case.

We have calculated two cases which neglect quantum corrections.
One corresponds to the small $\lm$, $\lm\sim O(1)$, and the other
the large one, $\lm\sim O(100)$.
The latter is the order used in the low-energy effective model of the
$\pi, \sigma$ mesons.
The results are as follows.
Without the SS diagram, if the coupling constant $\lm$ is small,
the weak first order phase transition occurs.
We see that in this case the order becomes second by including the SS diagram,
while in the strong coupling case the order does not change.
So the contribution of the SS diagram in this approximation changes the
order of the phase transition from the first to the second if the original 
first order is rather weak.

Chiku and Hatsuda calculated the SS diagram in the optimized perturbation
theory\cite{Chi98}.
They include the resummed mass parameter in the lagrangian from the
beginning and renormalize it first with the approximation
that the mass parameter is
momentum independent constant factor.
We think that their approximation is at the same level as
ours used above.
Nevertheless they suggested that the SS diagram may change
the order of the phase transition from the first to
the second\cite{Chi99}.

 Nachbagauer also calculated the SS diagram in the one-component
$\p^4$ theory in the CJT action\cite{Nac95}.
He calculated the self-energy in a certain approximation in order to
get rid of the UV-divergence and
evaluated the Green function in the Pad\'e approximation.
He also pointed out that this diagram is important near the critical 
temperature, although he does not evaluate the order of the transition.

  In conclusion, it seems to be necessary to include the effects of the 
higher-order diagrams, 
if we discuss the behavior near the critical temperature.
Further investigation is needed in the future.

\section{Summary and Conclusion}
  \label{chp:summ}

  We have studied the $O(N)$ symmetric linear sigma model at finite
temperature in the Hartree-Fock approximation using the CJT effective
potential.
This model has been considered as the low-energy effective model of the
low-energy mesons, i.e., the sigma meson and the pions.
It has so far been understood that the NG theorem is not
satisfied, or the pions are massive at finite temperature in the
CJT formalism unless the large $N$ limit is taken.
We showed that this is caused by an improper derivation of the 
effective potential, such as an $O(N)$ violating form for the solution
of the SD-equation for the propagator.
If $O(N)$ symmetric form is chosen and one defines the meson masses as the
second derivatives of the effective potential, the NG theorem is
always satisfied even at finite values of $N$.
This is also satisfied at any order of the loop expansion of $\Gamma_2$,
which is the contribution of the two-particle irreducible vacuum graphs.

  We have also used a renormalization prescription of the CJT 
effective potential
from the analogy to renormalization of the auxiliary field.
This renormalization works apparently both in the symmetry broken 
and restored phases
while the conventional renormalization works only in the symmetry 
restored phase.

  When one is interested in effects of 
the thermal contributions at finite temperature,
quantum corrections are often neglected.
The classical equation of motions are satisfied at $T=0$ in this case
and the thermal effects are added as temperature increases.
We have solved the gap equation for the sigma condensate, the
SD-equations and the meson masses numerically in this approximation.
As a result the first order phase transition occurs, because 
this corresponds to the mean field approach.
Related to this approximation, we encounter a problem in calculating
the sigma meson mass in the chiral limit because of the infrared
singularity.
This difficulty is removed if one includes the quantum corrections
because the solutions of the SD-equations which cause the infrared
singularity remain finite in the chiral limit at $T=0$.
So consistent treatments including the quantum corrections which
is discussed in Sec. \ref{sec:renorm} are needed to obtain more
realistic results.
In Subsec. \ref{subsub:qc}, based on the above notion,
we incorporated the quantum corrections.
As a result the infrared singularity disappeared and the behavior apparently
became better for both the chiral limit and the finite pion mass.

  The resulting phase transition in $O(4)$ linear sigma model
is of the first order which is consistent with other mean field approaches.
However the second order phase transition is reported in this model 
in the renormalization group equation
analyses.
In fact near the critical temperature other loop contributions may become
significant and we showed the simple estimation of the effects of the
setting-sun diagram in the CJT approach.
As a result the order of the phase transition can change with this inclusion.
Therefore further investigation is needed if one studies
the order of the phase transition more rigorously.\\

\appendix
{\Large \bf Appendix}
\section{The $O(N)$ linear sigma model with auxiliary fields}
We applied the technique of the renormalization of auxiliary fields
for the CJT approach in the main text.
In this appendix we analyze the $O(N)$ linear sigma
model at finite temperature by using auxiliary fields
in order to show a similarity with the CJT approach.
We also point out that if we analyze the $O(N)$ linear sigma model
at finite temperature with the auxiliary fields from the beginning,
some difficulties associated with the NG theorem arise again.
Such an approach was carried out in ref.\cite{Bil99},
where one auxiliary field was introduced.
However this causes the violation of the NG theorem unless some 1-loop
terms are added in the sigma and pion self-energy.
We require a more general form of auxiliary fields which do not break the
$O(N)$ symmetry in order to relate it to the CJT approach.

 For the original lagrangian, after shifting the fields
$\p^a \to \p^a + \ppc^a$, we introduce an auxiliary field with an $N\times N$
symmetric matrix.
\begin{eqnarray}
  \lefteqn{\exp\left\{ -\int d^4x \frac{\lambda_0}{6N}(\phi^2)^2 \right\} }
  \nonumber \\
  &=&
   \int {\cal D}\chi \exp\left\{-\int d^4x\left[\frac{\lambda_0}{6N}(\phi^2)^2
  -\frac{3N}{8\lambda_0(N+8)}\left( \chi^{ab}-\frac{2\lambda_0}{3N}\phi^2
  \delta^{ab}-\frac{4\lambda_0}{3N}\phi^a\phi^b\right)^2 \right] \right\} \nn
\end{eqnarray}
We take the saddle point approximation for $\chi^{ab}$, i.e.,
$\chi^{ab}\approx \chi_c^{ab}$, and then the effective potential
at the 1-loop order is obtained.
\begin{eqnarray}
  V(\phi_c,\chi_c) &=& \frac{m^2}{2}\phi_c^2 + 
  \frac{\lambda}{6N}(\phi_c^2)^2 - \frac{3N}{8\lambda(N+8)}\chi_c^2 
  \nonumber \\
  & & +\frac{1}{2} \intbk \ln \det \left\{ \left( k^2+m^2+
  \frac{2\lambda}{3N}\phi_c^2 + \frac{\chi_c^{cc}}{N+8} \right) \delta^{ab}
  +\frac{4\lambda}{3N}\phi_c^a \phi_c^b + \frac{2}{N+8} \chi_c^{ab} \right\}
  \nonumber \\ &&
  \label{eq:effa}
\end{eqnarray}
From the extremal conditions for $\chi_c^{ab}$ and $\ppc^a$,
we obtain the following equations.
\begin{equation}
  M_1^2 = m^2 + \frac{2\lambda}{3N} + \frac{2\lambda}{3}
  \frac{1}{N+8}\intbk \left\{ \frac{N+3}{k^2+M_1^2} + 
  \frac{(N+4)/N}{k^2+M_2^2} \right\}
  \label{eq:sd1a}
\end{equation}
\begin{equation}
  M_2^2 = m^2 + \frac{2\lambda}{N} + \frac{2\lambda}{3}
  \frac{1}{N} \intbk \left\{ \frac{(N+4)(N-1)/(N+8)}{k^2+M_1^2} + 
  \frac{1}{k^2+M_2^2} \right\}
  \label{eq:sd2a}
\end{equation}
\begin{equation}
  \phi_c^2 \left\{m^2 + \frac{2\lambda}{3N}\phi_c^2 + 
  \frac{2\lambda}{N}\intbk \frac{1}{k^2+M_2^2} +
  \frac{2\lambda(N-1)}{3N}\intbk \frac{1}{k^2+M_1^2} \right\} = 0
  \label{eq:gapa}
\end{equation}
where we defined $\alpha$ and $\beta$ by
$\chi_c^{ab} = \alpha \delta^{ab} + \beta \frac{\phi_c^a\phi_c^b}{\phi_c^2}$
without loss of generality of $\chi_c^{ab}$ and put
\begin{equation}
  M_1^2 = m^2 + \frac{2\lambda}{3N}\phi_c^2
   + \frac{(N+2)\alpha+\beta}{N+8}
\end{equation}
\begin{equation}
  M_2^2 = m^2 + \frac{2\lambda}{N}\phi_c^2
   + \frac{(N+2)\alpha+3\beta}{N+8}
\end{equation}
The equations (\ref{eq:sd1a}), (\ref{eq:sd2a}) and (\ref{eq:gapa})
are similar to eqs.(\ref{eq:sds}), (\ref{eq:sdp}) 
and (\ref{eq:gap}) in the CJT approach, respectively.
In fact the renormalized form of the effective potential (\ref{eq:effa})
can be obtained by requiring renormalization conditions for $M_1$ and
$M_2$ at zero temperature, 
which means the renormalization of the auxiliary fields.
Our renormaliztion prescription in the CJT approach is based on this
idea.

A difficulty occurs, however, when we consider the case of finite
temperature.
Our numerical calculation shows that
the equations (\ref{eq:sd1a}), (\ref{eq:sd2a}) and (\ref{eq:gapa})
do not have any physical solutions at finite temperature in contrast 
to the CJT approach.
This means that the summation of the super-daisy diagrams does not
work well and we may require some additional terms to recover it
as in ref.\cite{Bil99}.
The CJT approach, on the other hand, does not have such a problem,
though the forms of the equations are similar.
We think that the CJT case corresponds to special auxiliary fields 
other than those of this section, but the relations between them
remain as the future problem.



\clearpage

\begin{table}[t]
\begin{center}
\caption{Initial Parameters for $T=0$}
$m_\pi = 0$ MeV
\begin{tabular}{|c|c||c|c|c|c|c|} 
\hline
$m_\sigma\ [{\rm MeV}]$ & 
$\mu\ [{\rm MeV}]$ & $\lambda$ & $M_X\ [{\rm MeV}]$  & $M_{XY}\ 
[{\rm MeV}]$  & $m^2\ [{\rm MeV}^2]$ &
$T'_c$ [MeV] \\ \hline \hline
600 & 320   & 90.2 & 41.5 & 510 & -122375 & 128  \\
600 & 400   & 102  & 119  & 542 & -94208  & 105  \\
600 & 500   & 122  & 183  & 594 & -47679  & 68.4 \\
600 & 567   & 139  & 223  & 632 & 0       & 0    \\ 
600 & TREE  & 125  & 0    & 600 & -180000 & 132 \\ \hline \hline
1000 & 410   & 156 & 29  & 670 & -218034 & 130  \\
1000 & 500   & 177 & 156 & 715 & -128224 & 93.2 \\
1000 & 599   & 208 & 239 & 774 & 0       & 0    \\ 
1000 & TREE  & 347 & 0   & 1000& -500000 & 132  \\ \hline
\end{tabular}
  \label{tab:ini1}
\end{center}
\end{table}

\begin{table}[b]
                     \label{tab:init2}
\begin{center}
\caption{Initial Parameters for $T=0$ (various values of $\mu$)}
$m_\pi=138$ MeV \\
\begin{tabular}{|c|c||c|c|c|c|} 
\hline
$m_\sigma\ [{\rm MeV}]$ & 
$\mu\ [{\rm MeV}]$ & $\lambda$ & $M_X\ [{\rm MeV}]$  & $M_{XY}\ 
[{\rm MeV}]$  & $m^2\ [{\rm MeV}^2]$  \\ \hline \hline
600 & 220   & 66.9 & 55.9 & 460 & -96324 \\
600 & 300   & 77.8 & 114  & 493 & -83362 \\
600 & 400   & 94.7 & 165  & 540 & -59939 \\
600 & 520   & 119  & 223  & 603 & -5633  \\ 
600 & TREE  & 118  & 138  & 600 & -151434  \\ \hline \hline
1000 & 380   & 142 & 54.8 & 655 & -197043 \\
1000 & 400   & 146 & 87.5 & 664 & -180329 \\
1000 & 500   & 172 & 184  & 718 & -95776  \\
1000 & 580   & 196 & 245  & 765 & -106    \\ 
1000 & TREE  & 340 & 138  & 1000& -471434 \\ \hline
\end{tabular}
\end{center}
\end{table}
\begin{table}[h]
\begin{center}
\caption{Initial Parameters for $T=0$ (various pion masses)}
$m_\sigma = 600$ MeV, \qquad $\mu=320$ MeV \\
\begin{tabular}{|c||c|c|c|c|} 
\hline
$m_\pi\ [{\rm MeV}]$ & $\lambda$ & $M_X\ [{\rm MeV}]$  & $M_{XY}\ 
[{\rm MeV}]$  & $m^2\ [{\rm MeV}^2]$  \\ \hline \hline
0     & 90.2 & 41.5 & 510 & -122375  \\
20    & 89.8 & 44.7 & 509 & -121078  \\
60    & 87.8 & 64.8 & 507 & -112237  \\
100   & 84.6 & 93.5 & 504 & -97658 \\
138   & 80.9 & 125  & 502 & -79760 \\ \hline
\end{tabular}
  \label{tab:ini3}
\end{center}
\end{table}

\clearpage




\begin{thebibliography}{99}
\bibitem{Kir72} D. A. Kirzhnits and A. D. Linde, Phys. Lett. {\bf B 42},
471 (1972).

\bibitem{Wei74} S. Weinberg, Phys. Rev. {\bf D 9}, 3357 (1974).

\bibitem{Dol74} L. Dolan and R. Jackiw, Phys. Rev. {\bf D 9}, 3320 (1974).

\bibitem{Kir76} D. A. Kirzhnits and A. D. Linde, Ann. Phys. {\bf 101},
195 (1976).

\bibitem{Bay77} G. Baym and G. Grinstein, Phys. Rev. {\bf D 15}, 2897 (1977).

\bibitem{Lar86} \AA. Larsen, Z. Phys. C {\bf 33}, 291 (1986).

\bibitem{Roh98} H.-S. Roh and T. Matsui, Eur. Phys. J. A {\bf 1}, 205 (1998).

\bibitem{CJT74} J. M. Cornwall, R. Jackiw and E. Tomboulis,
Phys. Rev. {\bf D 15}, 2428 (1974).

\bibitem{Ame93} G. Amelino-Camelia and S.-Y. Pi, Phys. Rev. {\bf D 47}, 2356
(1993).

\bibitem{Chi98} S. Chiku and T. Hatsuda, Phys. Rev. {\bf D 57}, R6 (1998);
{\it ibid.} {\bf D 58}, 076001 (1998).

\bibitem{Chi99} S. Chiku and T. Hatsuda, {\it private communication}.

\bibitem{Ame97} G. Amelino-Camelia, Phys. Lett. {\bf B 407}, 268 (1997).

\bibitem{Pet98} N. Petropoulos, J. Phys. G {\bf 25},2225 (1999).

\bibitem{CJP74} J. M. Cornwall, R. Jackiw, and H. D. Politzer, Phys. Rev.
{\bf D 10}, 2491 (1974).

\bibitem{Kob75} M. Kobayashi and T. Kugo, Prog. Theor. Phys. {\bf 54},
 1537 (1975).

\bibitem{Lee60} T. D. Lee and C. N. Yang, Phys. Rev. {\bf 117}, 22
(1960).

\bibitem{Dah67} H. D. Dahmen and G. Jona-Lasinio, Nuovo Cimento A 
{\bf 52}, 807 (1967).

\bibitem{Oko96} A. Okopi\'nska, Phys. Lett. {\bf B 375}, 213 (1996).

\bibitem{Abb76} L. Abbott, J. S. Kang and H. J. Schnitzer, Phys. Rev.
{\bf D 13}, 2212 (1976).

\bibitem{Len99} J.T. Lenaghan and D.H. Rischke,
{\it nucl-th/9901049}.

\bibitem{PDG}   Particle Data Group, Eur. Phys. J. {\bf C} 3, 1  (1998).

\bibitem{Nai98b} K. Naito, K. Yoshida, Y. Nemoto, M. Oka and M. Takizawa,
Phys. Rev. {\bf C 59}, 1722 (1999).

\bibitem{Ga84} J. Gasser and H. Leutwyler, Ann. Phys. {\bf 158}, 142
(1984).

\bibitem{det87}  C.E. DeTar and J. Kogut, Phys. Rev. Lett. {\bf 59},
399 (1987);
S. Gottlieb et al, Phys. Rev. Lett. {\bf 59},1881 (1987);
Y. Koike, M. Fukugita and A. Ukawa, Phys. Lett. {\bf B 213},497 (1988);
K.M. Bitar et al, Phys. Rev. {\bf D 43},302 (1991);
K.D. Born et al, Phys. Rev. Lett. {\bf 67},302 (1992).

\bibitem{shy} T. Schaefer and E.V. Shuryak, Phys. Lett. {\bf B 356},
147 (1995).

\bibitem{Raj93} K. Rajagopal and F. Wilczek, Nucl. Phys. {\bf B 399},
395 (1993);
H. Nakkagawa and H. Yokota, Mod. Phys. Lett. A 11, 2259 (1996);
J. Berges, D.-U. Jungnickel and C. Wetterich, Phys. Rev. {\bf D 59},
034010 (1999);
T. Umekawa, K. Naito and M. Oka, {\it hep-ph/9905502}.

\bibitem{Ogu98} K. Ogure and J. Sato, Phys. Rev. {\bf D 58},
085010 (1998).

\bibitem{Arn93} P. Arnold and O. Espinosa, Phys. Rev. {\bf D 47},
3546 (1993); {\it ibid.} {\bf D 50}, 6662 (1994) (E).

\bibitem{Nac95} H. Nachbagauer, Z. Phys. C {\bf 67}, 641 (1995).

\bibitem{Bil99} N. Bili\'c and H. Nikoli\'c, Eur. Phys. J. C 6,
515 (1999).

\end{thebibliography}
\end{document}